\documentclass{aastex61}

\newcommand\aastex{AAS\TeX}

\received{xxx}
\revised{xxx}
\accepted{xxx}
\submitjournal{ApJ}

\shorttitle{\aastex\ sample article}
\shortauthors{Ishii et al.}
\begin{document}

\title{Free Neutron Ejection from Shock Breakout in Binary Neutron Star Mergers}

\correspondingauthor{Ayako Ishii}
\email{ishii@resceu.s.u-tokyo.ac.jp}

%\author[0000-0002-0786-7307]{Greg J. Schwarz}
%\affil{American Astronomical Society \\
%2000 Florida Ave., NW, Suite 300 \\
%Washington, DC 20009-1231, USA}

\author{Ayako Ishii}
\affiliation{Research Center for the Early Universe, Graduate School of Science, University of Tokyo, Bunkyo-ku, Tokyo 113-0033, Japan}
%\collaboration{(AAS Journals Data Scientists collaboration)}

\author{Toshikazu Shigeyama}
\affiliation{Research Center for the Early Universe, Graduate School of Science, University of Tokyo, Bunkyo-ku, Tokyo 113-0033, Japan}
%\collaboration{(LaTeX collaboration)}

\author{Masaomi Tanaka}
\affiliation{Astronomical Institute, Tohoku University, Aoba, Sendai 980-8578, Japan}
\affiliation{National Astronomical Observatory of Japan, Osawa, Mitaka, Tokyo 181-8588, Japan}
%\nocollaboration

\begin{abstract}
Merging neutron stars generate shock waves that disintegrate heavy nuclei into nucleons especially in the outer envelope. 
It is expected that some of these neutrons having avoided capturing positrons, remain as free neutrons even 
after the disappearance of electron--positron pairs.
To investigate how many free neutrons can be ejected from merging neutron stars,
we performed special-relativistic Lagrangian hydrodynamics computations 
with simplified models of this phenomenon in which a spherically symmetric shock wave propagates 
in the hydrostatic envelope and emerges from the surface. 
%%%% from Tanaka comment
%We introduce two parameters (the size of the merging neutron stars and the energy involved in the shock waves) 
%to describe the phenomenon. 
%As a result of a systematic study for a broad range of these parameters, it is found that free neutrons with a mass of $10^{-7}$ to $10^{-6}\,M_\odot$ remain in the ejecta. 
We systematically study a wide parameter space of the size of the merging neutron stars and the energy involved in the shock waves. 
As a result, it is found that the mass of remaining free neutrons is $10^{-7}$ to $10^{-6}\ M_{\odot}$, 
which is smaller than the previously expected mass by more than two orders of magnitude. 
%This is due to the p(n, $\gamma$)d reactions, which were not taken into account in previous studies.
%%%%
%There is a preferred energy of the order of $10^{48}$ erg for each size of the envelope 
There is a preferred energy of the order of $10^{48}$ erg 
that yields the maximum amount of free neutrons for large sizes of the envelope. 
%We briefly discuss the emission from the free neutron layer and estimate the peak luminosity in the optical band to be 
%about $6\times10^{38}$ erg s$^{-1}(M_{\rm n}/10^{-6}\,M_\odot)$ at 8,000 s after the merger.
We briefly discuss the emission from the free neutron layer and estimate the luminosity in the optical band to be 
about $7\times 10^{41}$ erg s$^{-1}(M_{\rm n}/10^{-6}\, M_\odot)$ in $\sim 30$ minutes after the merger.
\end{abstract}

%\keywords{binary neutron star merger --- shock breakout ---
%relativistic hydrodynamics}

\section{Introduction} \label{sec:intro}

Gravitational waves (GW170817) from a binary neutron star merger (NSM) were detected by LIGO on 2017 August 17 UT\citep{2017PhRvL.119p1101A}.  
%%%% from Tanaka comment
%The electromagnetic wave counterpart was also detected by Fermi \citep{2017arXiv171005450F, 2017ApJ...848L..14G} INTEGRAL \citep{2017ApJ...848L..13A, 2017ApJ...848L..15S},  
%Swift and NuSTAR \citep{2017Sci...358.1565E}, Chandra \citep{2017ApJ...848L..25H}, and other radio and optical telescopes\citep{2017ApJ...848L..12A, 2017Sci...358.1574S, 2017Natur.551...71T, 
%2017Sci...358.1583K, 2017Sci...358.1559K, 2017Sci...358.1570D, 2017ApJ...848L..18N, 2017ApJ...848L..16S, 2017ApJ...850L...1L, 2017arXiv171005462H, 2017ApJ...848L..33A, 2017Natur.551...64A, 
%2017ApJ...848L..29D, 2017PASA...34...69A, 2017ApJ...848L..32M, 2017ApJ...848L..24V, 2017Natur.551...67P, 2017Sci...358.1579H, 2017ApJ...848L..21A, 2017ApJ...850L..21K}. 
The electromagnetic wave counterpart was also detected over the wide wavelength range \citep{2017arXiv171005450F, 2017ApJ...848L..14G, 2017ApJ...848L..13A, 2017ApJ...848L..15S,  
2017Sci...358.1565E, 2017ApJ...848L..25H, 2017ApJ...848L..12A, 2017Sci...358.1574S, 2017Natur.551...71T, 
2017Sci...358.1583K, 2017Sci...358.1559K, 2017Sci...358.1570D, 2017ApJ...848L..18N, 2017ApJ...848L..16S, 2017ApJ...850L...1L, 2017arXiv171005462H, 2017ApJ...848L..33A, 2017Natur.551...64A, 
2017ApJ...848L..29D, 2017PASA...34...69A, 2017ApJ...848L..32M, 2017ApJ...848L..24V, 2017Natur.551...67P, 
2017Natur.551...75S, 2017Sci...358.1579H, 2017ApJ...848L..21A, 2017ApJ...850L..21K, 2017arXiv171005865T}. 
The optical counterpart (SSS17a) was discovered by the Swope Supernova Survey 10.9 hr after GW170817\citep{2017Sci...358.1556C}.  
This observation singled out the host galaxy as an S0 galaxy NGC 4993. The distance to this galaxy was estimated to be $\sim$40 Mpc.  
%\citet{2017arXiv171005865T} confirmed that this is the only optical transient counterpart of GW 170817 in the error box of the localization from the combined observations of LIGO and Virgo.

The following optical and near-infrared observations \citep[e.g.][]{2017ApJ...848L..17C,2017ApJ...848L..19C, 2017PASJ...69..101U, 
2017Sci...358.1556C, 2017ApJ...848L..27T} have revealed that the emission in optical bands decays in a couple of days, 
while the emission 
in near-infrared bands lasted for about 10 days. 
The optical counterpart is thought to originate from matter ejected from the NSM and is referred to 
as a kilonova or macronova\citep{1998ApJ...507L..59L, 2005astro.ph.10256K, 2010MNRAS.406.2650M}. 
The ejecta are composed of nascent r-process elements, most of which are radioactive and thus work as a heat source. 
Theoretical models of kilonovae \citep{2013ApJ...774...25K,2013ApJ...775...18B, 2013ApJ...775..113T, 2017arXiv170809101T, 2017PASJ...69..102T,2017Natur.551...80K} have shown that  the ejecta that include a significant amount of lanthanide elements are much more opaque for optical photons  than the ejecta composed of Fe. 
As a consequence, it is predicted that a kilonova is brighter in near the infrared bands than in the optical, 
and can keep its brightness for about 10 days, depending on the ejecta mass. 
The features of the near infrared emission of SSS17a, which are consistent with predictions, 
enable estimations of the ejecta mass ($\sim 0.03\,M_\odot$) 
and the mean velocity  ($\sim0.1c$, where $c$ denotes the speed of light; \citet{2017PASJ...69..102T}). 
%On the other hand, the observed bright optical emission and its fast decay may suggest another component of the ejecta with a Lanthanide free composition \citep{2017Natur.551...80K}, though it is still controvertial \citep[see][]{2017PASJ...69..102T}. 

The follow-up observations started in 10.9 hr after the merger event. 
If the earlier emission from the NSM had been detected, it would have provided us with a different kind of information \citep{2018arXiv180202164A}.
For example, it was suggested by smoothed particle hydrodynamics (SPH) simulations that the outermost layers 
in the ejecta are composed of free neutrons with a mass of the order of $10^{-4}\,M_\odot$ 
\citep{2013ApJ...773...78B,2015MNRAS.448..541J,2015MNRAS.452.3894G}. 
The layer expands at a speed higher than 0.1$c$. 
If this is true, the free neutron layer might contribute the earlier emission 
because free neutrons decay on a time scale of $\sim$10 minutes \citep{2015MNRAS.446.1115M}. 
Another interesting possibility is that the emission from the cocoon may be produced by a jet crossing the ejecta \citep{2018MNRAS.473..576G, 2017Sci...358.1559K}.

\begin{figure}[t]
\begin{center}
\plotone{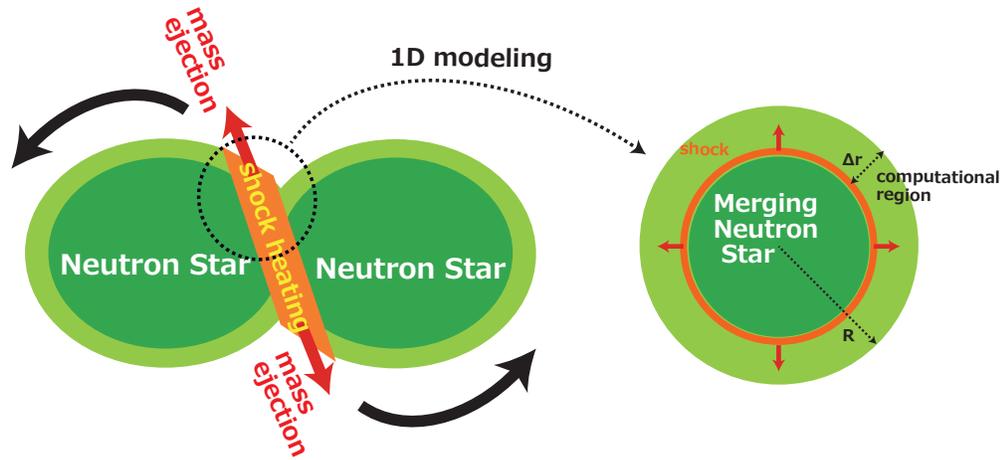}
\vspace{5mm}
\caption{Schematic picture of the binary NSM in reference to \citet{2014MNRAS.437L...6K}. 
	     Shock waves are generated in coalescing double-neutron stars and propagate into the tidally disrupted matter.
	     Some amount of material is ejected due to the shock heating.
	     The thickness $\Delta r$ of the outermost layer corresponds to the computational region 
	     and $R$ is the radius of the surface.}
\label{fig:NS-merger}
\end{center}
\end{figure}

%%%% from Tanaka comment
%Unfortunately, the existence of the free neutron layer was explored by a small number of SPH particles in their simulations. 
The emission powered by decays of free neutrons is not fully understood. 
The existence of the free neutron layer was explored using a small number of SPH particles 
in their simulations \citep{2015MNRAS.446.1115M}. 
We need to confirm if this is the case or we need to know under what conditions the free neutron layer can be ejected. 
NSMs have also been investigated using grid-based simulations. 
For example, \citet{2011PhRvL.107e1102S} and \citet{2012PhRvD..86f4032P} 
systematically studied NSMs by performing grid-based simulations and showed 
that shock waves broke out of matter near the contact surface of the two neutron stars (see Figure~\ref{fig:NS-merger}) 
though their simulations could not resolve matter with a mass down to $10^{-4}\,M_\odot$. 
\citet{2014MNRAS.437L...6K} discussed the omnidirectional emission from this kind of shock breakout. 

In this paper, we anticipate that the shock heating might disintegrate matter into free neutrons 
and protons and that the resultant layer containing free neutrons contributes to the early emission. 
To investigate this possibility, we utilize simplified spherically symmetric models 
in which we calculate the propagation of a  shock wave generated deep inside the surface layer 
with a density distribution following a power-law function of the distance to the surface.
 
The paper is organized as follows. 
In Section 2, we describe the setup of our simplified models in detail and introduce the numerical scheme we use 
to calculate the shock propagation and the subsequent phenomena. 
In Section 3, we present our results. Finally, in Section 4 we summarize the conditions for realizing a free neutron layer 
and discuss emission from this layer.

%・observation of gravitational wave from NS merger
%・ele-mag counterpart
%・early phase emission is not unveiled
%・theoretical work of early emission
%・free neutron precursor is focused
%・free neutron can be survive in SPH
%・How about grid-base simulation?
%・/fig neutron star ejecta
%・relativistic acceleration in Shock breakout (dynamical ejecta not equal wind)
%%%% if there is time...
%・/reaction rate of various process (fig?)
%%%%

%\section{Numerical modeling} 

\section{Methods} \label{sec:model}

%binary neutron star > BNS?
In a real NSM event, matter is ejected aspherically, so simulations for this event have been performed 
with numerical hydrodynamics codes fully incorporating multidimensionality. 
On the other hand, these comprehensive numerical studies may not be able to resolve the detailed structures close to the surface, 
which can affect emission in the early phase of the event. 
The free neutron layer mentioned above is one example. 
Here, we employ an alternative approach by simplifying the situation as depicted in Figure \ref{fig:NS-merger}, 
which is the strategy taken by \citet{2014MNRAS.437L...6K}. 

\subsection{Initial setting of our model}\label{initial}
A spherical envelope in hydrostatic equilibrium is placed close to the surface with a mass density distribution $\rho(r) \propto (R - r)^3$, where $R$ and $r$ are the radius of the surface and the radial coordinate, respectively.  
\citet{2014MNRAS.437L...6K} used $R=15$ km for their fiducial value from results of numerical simulations. 
It is expected and confirmed that a smaller radius requires a stronger shock to eject the same amount of the envelope, 
thus leading to a higher density of positrons, and reducing the mass of free neutrons by the positron capture process. 
Therefore, we explore models with $R$ = 15 -- 30 km to search for optimal conditions for free neutrons. 
The thickness $\Delta r$ of  the outermost layer (corresponding to the computational region) is determined 
so that the total mass $M_{\rm env}$ involved in this region becomes  $10^{-3}\, M_\odot$. 
The resultant $\Delta r$ takes the value of a few $\times 10^{-2}R$ with $R$ = 15 -- 30 km.  
This value of  $M_{\rm env}$ is chosen to be significantly greater than the mass of ejecta to reduce the influence of the method of the  initial energy input to generate a shock wave on the amount of free neutrons. 
The gravitational energies of the envelope, less than $-10^{50}$ erg, guarantee that only a part of the envelope finally becomes unbound. 

%%%% if there is time
%shock propagate into tidal disrupted ejecta?
%%%%
The pressure $P(r)$ in this region is assumed to be dominated by degenerate ultra-relativistic electron gas  given by 
\begin{equation}
P(r)=\frac{(3\pi^2)^{1/3}\hbar c}{4}\left(\frac{Y_{\rm e}\rho(r)}{m_{\rm u}}\right)^{4/3}\equiv K\rho(r)^{4/3},
\end{equation}
where $\hbar$,  $Y_{\rm e}$, and $m_{\rm u}$ denote the Dirac constant, the electron fraction per nucleon, 
and the atomic mass unit, respectively. 
Here, we assume that $Y_{\rm e}$ has a constant value of 0.1 to mimic the beta equilibrium in cold dense matter. A shock wave is generated by increasing the thermal energy in the innermost region. 

\subsection{Hydrodynamics code} \label{sec:method}
To describe the shock propagation in the envelope, we have developed a special-relativistic Lagrangian hydrodynamics code based on the previous work introduced in \citet{2003rnh..book.....W}. 
Equations of Lagrangian hydrodynamics with the time $t$ and radial coordinates $r$ are written as follows:
\begin{eqnarray}
\frac{d D}{d t} + D\frac{\partial r^2 v_r}{r^2\partial r}&=& 0, \\
\frac{d E}{d t} + E\frac{\partial r^2 v_r}{r^2\partial r}+ P \left[ \frac{d \Gamma}{dt} + \Gamma\frac{\partial r^2 v_r}{r^2\partial r}\right] &=& 0, \\
\frac{d S_r}{d t} + S_r \frac{\partial r^2 v_r}{r^2\partial r}+ \frac{\partial P}{\partial r}+\frac{GM\rho}{r^2}&=& 0,
%\frac{d D}{d t} + D \nabla \cdot v^i &=& 0, \\
%\frac{d E}{d t} + E \nabla \cdot v^i + P \left[ \frac{\partial W}{\partial t} + W \nabla \cdot v^i + v^i \nabla W \right] &=& 0, \\
%\frac{d S_j}{d t} + S_j \nabla \cdot v^i + \frac{\partial P}{\partial x^j} &=& 0,
\end{eqnarray}
%where $v_r$, $W$, and $P$ are the radial velocity, the Lorentz factor, and the pressure, respectively. 
where $v_r$, $\Gamma$, $G$, and $M$ are the radial velocity, the Lorentz factor, the gravitational constant, and the mass of a merging object, respectively. 
The term $GM\rho/r^2$, expressing the gravitational force per volume, is added to the momentum equation. 
We have fixed the value of $M$ to 2 $M_\odot$ throughout the computations. 
Here, we have chosen a mass significantly smaller than the total mass of the merging object ($\sim2.8 M_{\odot}$) 
to partially take into account the effects of the centrifugal force on the surface gravity in our spherically symmetric model.
The Lorentz-contracted state variables $D$, $E$, and $S_r$ denote the coordinate baryon mass density, 
the coordinate internal energy density, and the coordinate momentum density, respectively, 
and can be expressed with the mass density $\rho$, the specific internal energy $\epsilon$ defined in the comoving frame, 
and $v_r$ as
\begin{eqnarray}
D &=& \Gamma \rho, \\
E &=& \Gamma \rho \epsilon, \\
S_r &=& (D + \gamma E) \Gamma v_r. 
\end{eqnarray}
%where $\rho$ and $\epsilon$ are the mass density and the internal energy density defined in the co-moving frame of the fluid, respectively.
We use the ultra-relativistic equation of state  $P = \rho \epsilon/3$ and thus $\gamma=4/3$. 
The finite difference approach was adopted with a leapfrog scheme and the artificial viscosity method was employed. 
The computation was performed in the geometrical units with $G = c = 1$.

This Lagrangian hydrodynamics code was validated through  solutions of some shock tube problems presented in \citet{2003LRR.....6....7M} 
and we have confirmed that the code reproduces the results of a known self-similar solution \citep{Sakurai1960} 
relevant to our present work, 
in which a shock wave propagates in a non-uniform medium through the medium-vacuum interface. 

We use 500 computational cells to cover the entire region. 
We initiate calculations by increasing the thermal energy in the innermost 10 cells to generate a strong shock wave.
We use the final kinetic energy $E_{\rm f}$ to distinguish models with different  inputs of the thermal energy inputs.
%Table 1 lists the actual values of these four(?) quantities  used in our calculations. 
%We impose the boundary condition that the pressure is equal to zero at the surface. 
The  velocity at the innermost computational cell is always set to be zero  to mimic the stiff outer crusts of the neutron stars.

The temperature $T$ after the passage of the shock is derived from the formula 
\begin{equation}
P = \frac{a_{\rm r} T^4}{3}+K\rho^{4/3}, 
\end{equation}
where $a_{\rm r} $ is the radiation constant ($= 7.56 \times 10^{-15}\ {\rm erg\ cm^{-3}\ K^{-4}}$). 
Here, the pressure is approximated by the sum of the radiation pressure and the degeneracy pressure of ultra-relativistic electron gas.

\subsection{Abundance of free neutrons} \label{sec:neutron}
We have calculated the temporal evolution of the mass fraction of free neutrons in each cell after the shock hits the cell. 
It is assumed that the shock wave immediately disintegrates nuclei of heavy elements to free nucleons due to high temperatures ($T>10^{10}$ K) at the shock front. 
%\begin{equation}\label{eq:beta}
%\frac{n_{\rm n}}{n_{\rm p}} = {\rm exp} \left( - \frac{m_n c^2- m_p c^2}{kT} \right), 
%\end{equation}
%where $n_{\rm n}$ and $n_{\rm p}$ denote the number density of neutrons and protons,  $m_{\rm n}$ and $m_{\rm p}$ denote the mass of neutron and proton, and $k$ denotes the Boltzmann constant.The difference between the rest mass energies of a neutron  and a proton $m_{\rm n}c^2- m_{\rm p}c^2$ is around 1.293 MeV.  Thus the resultant $X_{\rm n}$ is always smaller than 0.5, though neutron stars actually have neutron-rich material  at least before the shock passage.
Then, the positron capture process, $n + e^{+} \rightarrow p + \bar{\nu}_e$, decreases the number of these free neutrons, 
%{\bf while the electron captures} $p+e^-\rightarrow n+\nu_e$ {\bf produce free neutrons }.
while the electron capture process, $p+e^{-} \rightarrow n+\nu_e$, produces free neutrons. 
Since the timescale of the positron capture process $\tau_{+}$ is given in \citet{2015MNRAS.446.1115M} as 
\begin{equation}
\label{eq:posi_cap}
\tau_{+} \simeq 2.1 \left( \frac{kT}{{\rm MeV}} \right)^{-5}\ {\rm s}
\end{equation}
%The timescale is estimated to be $\sim 2.1 \times 10^{-5}$ s with $10^{11}$ K obtained in Section \ref{sec:shock} by Eq.~(\ref{eq:posi_cap}), 
%and it is larger than the ejecta cooling timescale; 
%Therefore free neutrons can  survive since ejecta is cooled before the positron capture process occurs. 
while the timescale of the electron capture process $\tau_{-}$ is given in \citet{1992STIN...9225163K}, 
we calculate the temporal variation of the mass fraction of free neutrons $X_{\rm n}$ by integrating
\begin{equation}
\label{eq:time_posi}
\frac{d X_{\rm n}}{dt} = - \frac{X_{\rm n}}{\Gamma \tau_{+} (T)}+\frac{(1-X_{\rm n})}{\Gamma \tau_{-} (T)}, 
\end{equation}
%with respect to the proper time $\tau$ in each cell to estimate how many free neutrons can be ejected.
with respect to the  time $t$ in the rest-frame of each cell to estimate how many free neutrons can be ejected, 
because the outermost layer expands with highly relativistic speeds. 
The initial value of $X_{\rm n}$ is set to be 0.9 according to the beta equilibrium of cold dense matter  (see section \ref{initial}).

After the temperature in each cell decreases down to $10^{10}$ K, we perform nuclear reaction network calculations using a code developed 
by one of the authors \citep{2010AIPC.1279..415S}, 
with the initial conditions (i.e., the abundances of protons and neutrons)  given by the above calculations 
to check how many free neutrons remaining at the temperature of $10^{10}$ K prevent from positron capture reactions 
and neutron capture reactions by heavier nuclides at lower temperatures. 
%Here it is assumed that the density homologously decreases proportional to $\tau^{-3}$ and that the temperature adiabatically decreases as $\tau^{-1}$, 
%where $\tau$ denotes the proper time. 
%We will show results of this calculation when the temperature in each cell decreases lower than $5\times10^7$ K in the next section. 

Here, neutrino cooling of the ejecta is neglected. 
The timescale for the e$^+$--e$^-$ pair annihilation ($\sim10^{-4}$ s) is longer than the expansion timescale even at the highest temperature of $\sim 10^{11}$ K 
in our computations \citep{1961AnPhy..16..321C}. 

\section{Results}
\subsection{Overview of shock breakouts} \label{sec:shock}
We have performed simulations for the 16 models listed in Table \ref{tab:table}: We have calculated models with four different values for the radii $R=15,\,20,\,25,$ and $30$  km.  
For each value of $R$, models with four different values of the final kinetic energies $E_{\rm f}$, i.e., $10^{47}, 10^{48}, 10^{49}$, and $10^{50}$ erg have been calculated.

\begin{figure}[t]
\plottwo{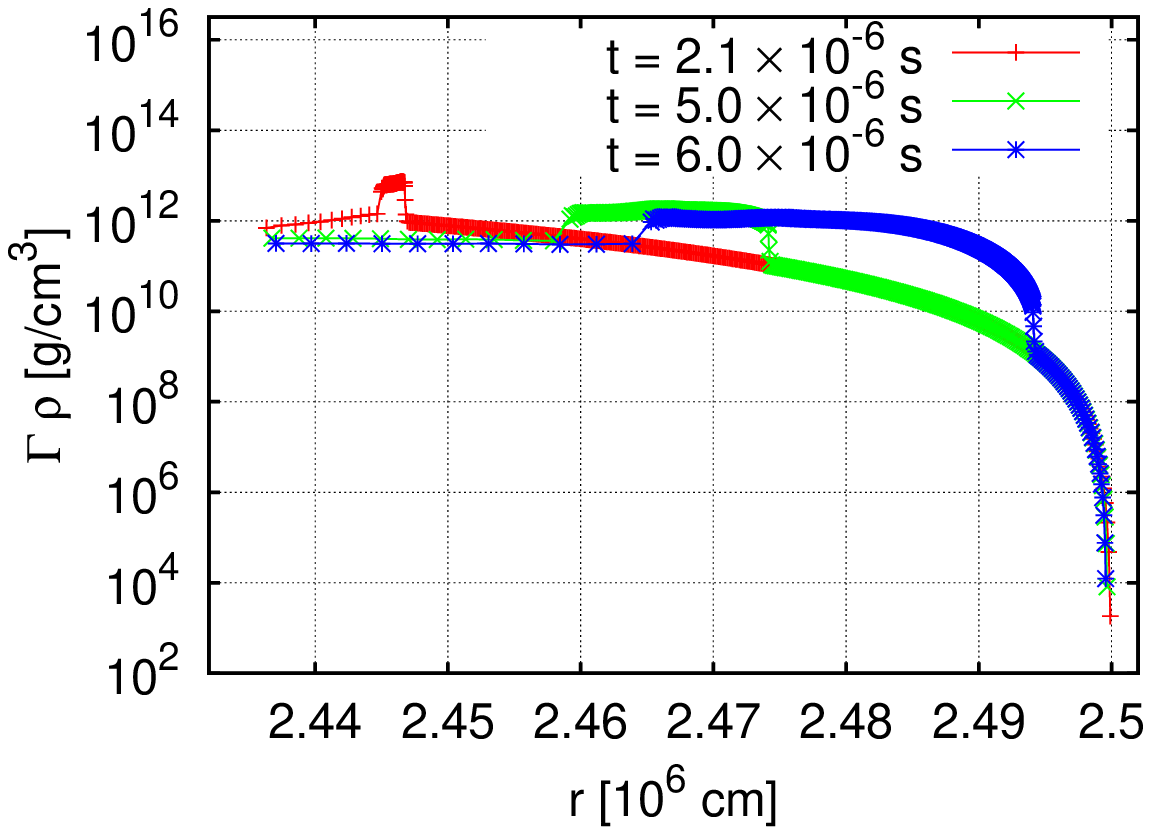}{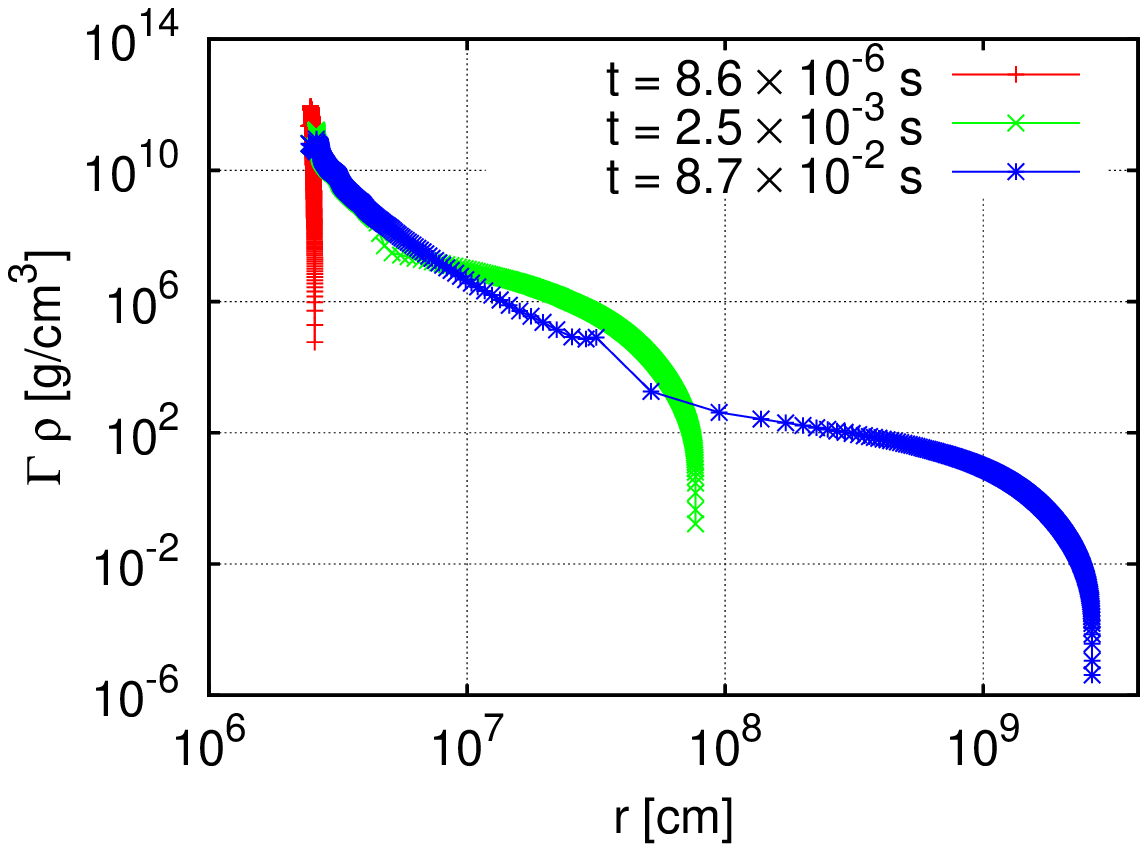}
%\vspace{10mm}
%\caption{Density distributions.}
%\label{fig:dens_distribution}
%\end{figure}
%
%\begin{figure}[t]
\plottwo{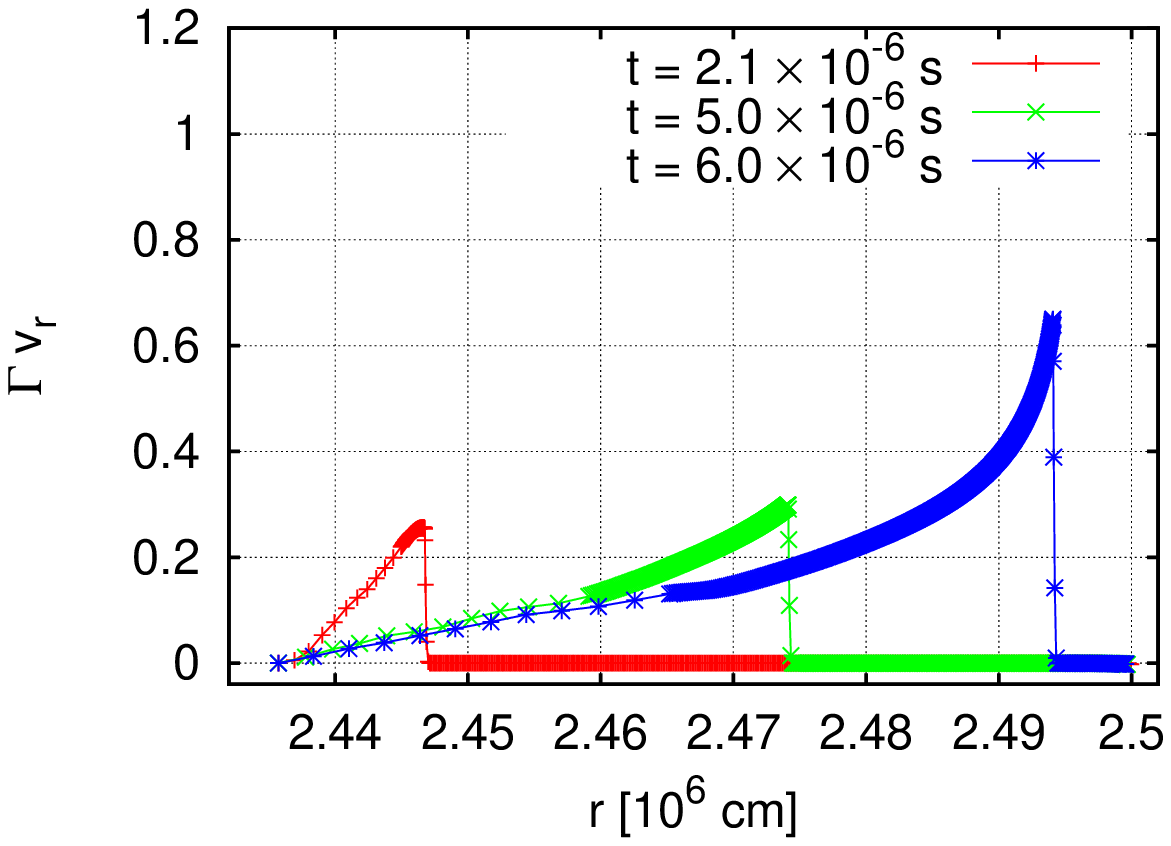}{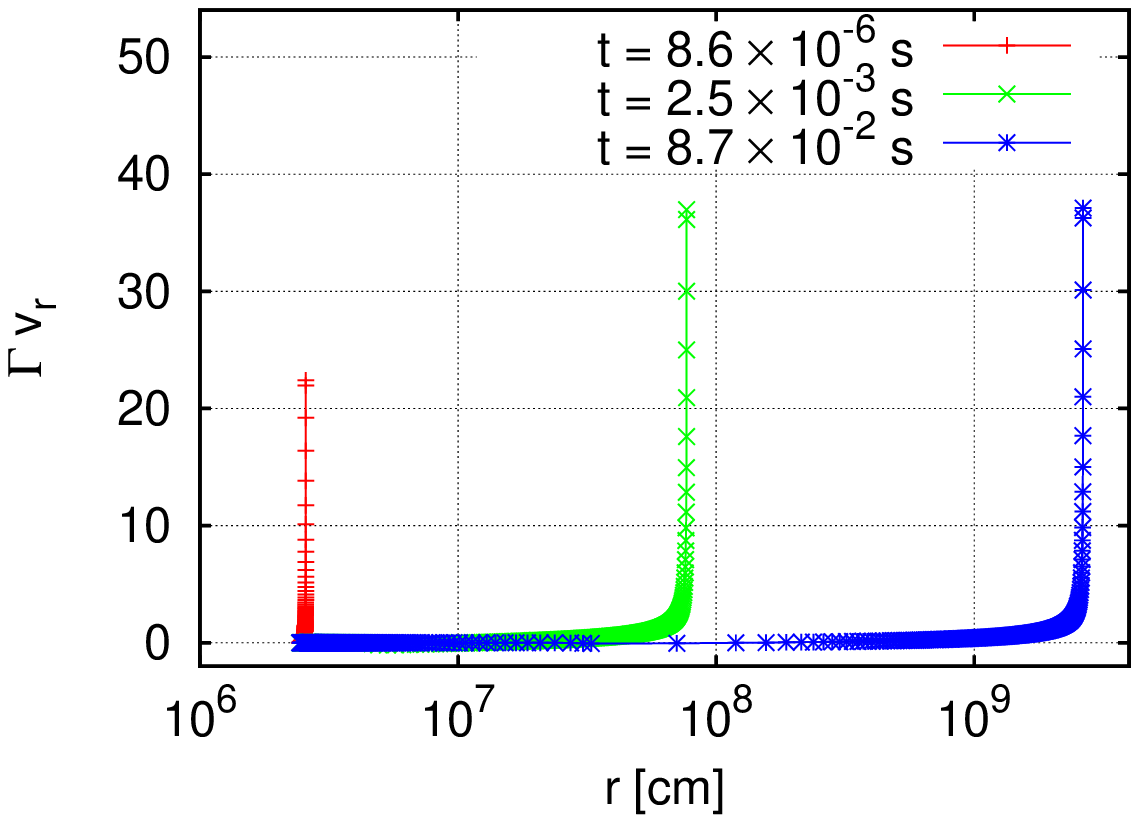}
%\vspace{10mm}
%\caption{Velocity distributions.}
%\label{fig:velo_distribution}
%\end{figure}
%
%\begin{figure}[t]
\plottwo{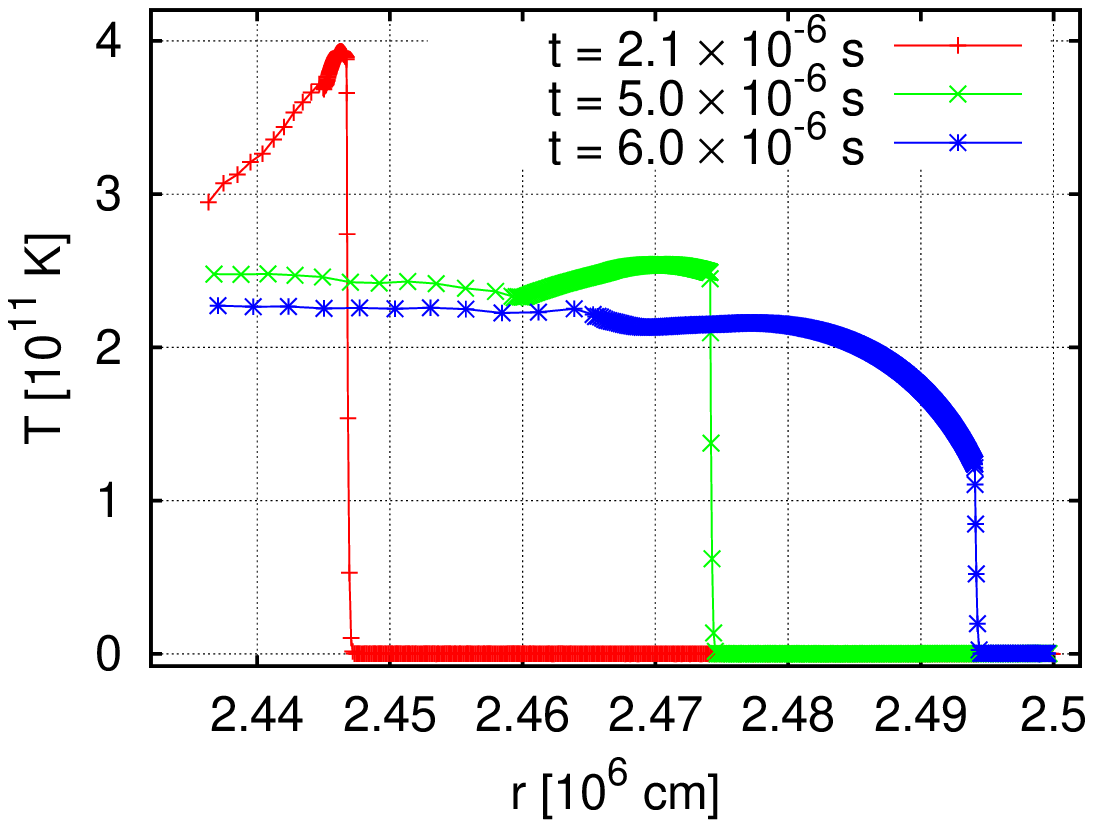}{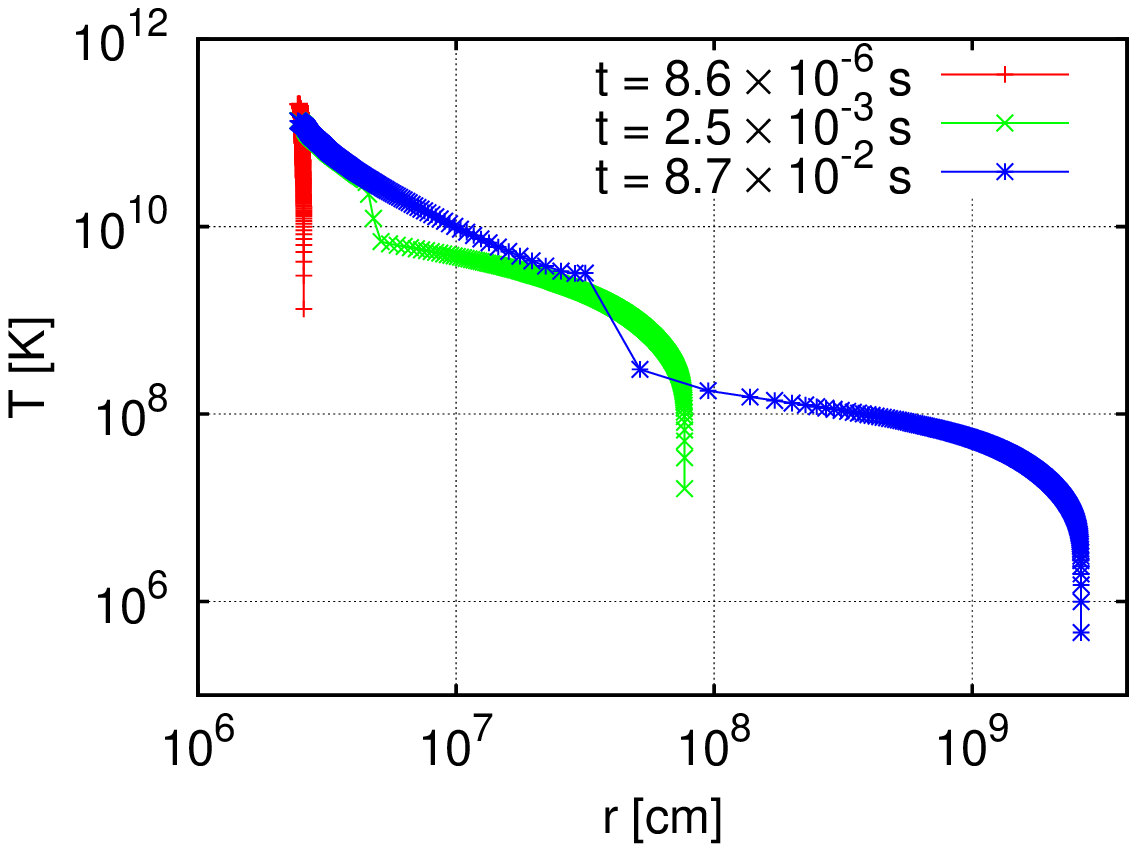}
\vspace{10mm}
%\caption{Temperature distributions.}
\caption{Temporal evolution of the density (top panels), the velocity (middle panels), and the temperature (bottom panels) profiles for a model with $R$ = 25 km and $E_{\rm f} = 10^{49}$ erg. 
	The left column shows the profiles before the shock breakout and the right column shows those after the shock breakout.}
\label{fig:dens_velo_temp}
\end{figure}

First, we present the results of a model with $R$ = 25 km and $E_{\rm f} = 10^{49}$ erg. 
Figure \ref{fig:dens_velo_temp} shows 
the temporal evolutions of the density ($\Gamma\rho$), the velocity ($\Gamma v_r$), and the temperature. 
Here, the left column shows the profiles before the shock breakout and the right one shows those after the shock breakout. 
%A strong shock wave propagates into the envelope before the shock breakout, and the density after the shock breakout keeps its functional form with respect to the distance from the surface as the Sakurai solutions\citep{Sakurai1960}.
The outermost region is accelerated to relativistic speeds (the middle right panel) 
and is consistent with \citet{2014MNRAS.437L...6K}. 
The temperature at the shock front always exceeds $10^{11}$ K (the bottom left panel). 
Actually, most of the ejecta experience a maximum temperature above $10^{10}$~K (see the right panel of Figure \ref{fig:velo_rho_temp}). 
At these temperatures, the most stable nuclei $^{56}$Fe are disintegrated to 13$\alpha+$4n and 
subsequently $\alpha$ nuclei are disintegrated to free nucleons on much shorter timescales than the light crossing time of the envelope, $\sim10^{-4}$ s. 
%Although it is unclear which elements exist in the outer layer before the merger, 
Thus, whatever elements exist before the merger, all the elements in the envelope will be disintegrated into nucleons after the shock passage because even the most stable nucleus $^{56}$Fe can be disintegrated by the shock. 
Then, a free neutron captures a positron on a time scale of $\sim 2.1 \times 10^{-5}$ s at $\sim10^{11}$ K (cf., eq.~(\ref{eq:posi_cap})), 
which is comparable to the timescale of the adiabatic cooling and might be longer, especially in the outer region where the shock cannot heat up the matter to such high temperatures. 
%Although a free neutron may react with a proton to produce a deuteron, 
%we have monitored the timescale of this reaction throughout the events and show the resultant distribution of mass fraction in the next section.

%\startlongtable
%\begin{deluxetable}{cccc}
\begin{deluxetable*}{cccc}[t]
\tablecaption{Ejecta masses and ejected free neutron masses for 16 different models. \label{tab:table}}
\tablehead{
\colhead{$R$} & \colhead{$E_{\rm f}$} & \colhead{$M_{\rm ej}$} & \colhead{$M_{\rm n}$}\\
%\colhead{} & \colhead{cost} & \colhead{charges\tablenotemark{b}} & \\
%\colhead{[km]} & \colhead{[erg]} & \colhead{[g]} & \colhead{[g]}
\colhead{(km)} & \colhead{(erg)} & \colhead{($M_{\odot}$)} & \colhead{($M_{\odot}$)}
}
%\colnumbers
\startdata
%15 & $10^{47}$ & $1.28 \times 10^{27}$ & $7.28 \times 10^{26}$ \\
%15 & $10^{48}$ & $1.18 \times 10^{28}$ & $8.93 \times 10^{26}$ \\
%15 & $10^{49}$ & $9.41 \times 10^{28}$ & $6.13 \times 10^{26}$ \\
%15 & $10^{50}$ & $6.08 \times 10^{29}$ & $3.92 \times 10^{26}$ \\
%15 & $10^{47}$ & $1.3 \times 10^{27}$ & $7.3 \times 10^{26}$ \\
%15 & $10^{48}$ & $1.2 \times 10^{28}$ & $8.9 \times 10^{26}$ \\
%15 & $10^{49}$ & $9.4 \times 10^{28}$ & $6.1 \times 10^{26}$ \\
%15 & $10^{50}$ & $6.1 \times 10^{29}$ & $3.9 \times 10^{26}$ \\
15 & $10^{47}$ & $6.4 \times 10^{-7}$ & $3.8 \times 10^{-7}$ \\
15 & $10^{48}$ & $6.0 \times 10^{-6}$ & $9.2 \times 10^{-7}$ \\
15 & $10^{49}$ & $4.7 \times 10^{-5}$ & $7.6 \times 10^{-7}$ \\
15 & $10^{50}$ & $3.0 \times 10^{-4}$ & $1.5 \times 10^{-6}$ \\
\hline
%20 & $10^{47}$ & $1.71 \times 10^{27}$ & $1.05 \times 10^{27}$ \\
%20 & $10^{48}$ & $1.56 \times 10^{28}$ & $2.16 \times 10^{27}$ \\
%20 & $10^{49}$ & $1.22 \times 10^{29}$ & $1.40 \times 10^{27}$ \\
%20 & $10^{50}$ & $7.78 \times 10^{29}$ & $8.95 \times 10^{26}$ \\
%20 & $10^{47}$ & $1.7 \times 10^{27}$ & $1.1 \times 10^{27}$ \\
%20 & $10^{48}$ & $1.6 \times 10^{28}$ & $2.2 \times 10^{27}$ \\
%20 & $10^{49}$ & $1.2 \times 10^{29}$ & $1.4 \times 10^{27}$ \\
%20 & $10^{50}$ & $7.8 \times 10^{29}$ & $9.0 \times 10^{26}$ \\
20 & $10^{47}$ & $8.6 \times 10^{-7}$ & $5.4 \times 10^{-7}$ \\
20 & $10^{48}$ & $7.9 \times 10^{-6}$ & $2.1 \times 10^{-6}$ \\
20 & $10^{49}$ & $6.1 \times 10^{-5}$ & $1.5 \times 10^{-6}$ \\
20 & $10^{50}$ & $3.9 \times 10^{-4}$ & $2.0 \times 10^{-6}$ \\
\hline
%25 & $10^{47}$ & $2.25 \times 10^{27}$ & $1.44 \times 10^{27}$ \\
%25 & $10^{48}$ & $1.98 \times 10^{28}$ & $4.30 \times 10^{27}$ \\
%25 & $10^{49}$ & $1.55 \times 10^{29}$ & $2.70 \times 10^{27}$ \\
%25 & $10^{50}$ & $9.39 \times 10^{29}$ & $1.67 \times 10^{27}$ \\
%25 & $10^{47}$ & $2.3 \times 10^{27}$ & $1.4 \times 10^{27}$ \\
%25 & $10^{48}$ & $2.0 \times 10^{28}$ & $4.3 \times 10^{27}$ \\
%25 & $10^{49}$ & $1.6 \times 10^{29}$ & $2.7 \times 10^{27}$ \\
%25 & $10^{50}$ & $9.4 \times 10^{29}$ & $1.7 \times 10^{27}$ \\
25 & $10^{47}$ & $1.1 \times 10^{-6}$ & $7.0 \times 10^{-7}$ \\
25 & $10^{48}$ & $1.0 \times 10^{-5}$ & $3.6 \times 10^{-6}$ \\
25 & $10^{49}$ & $7.4 \times 10^{-5}$ & $2.8 \times 10^{-6}$ \\
25 & $10^{50}$ & $4.6 \times 10^{-4}$ & $2.6 \times 10^{-6}$ \\
\hline
%30 & $10^{47}$ &  $2.67 \times 10^{27}$ & $1.75 \times 10^{27}$ \\
%30 & $10^{48}$ &  $2.36 \times 10^{28}$ & $7.60 \times 10^{27}$ \\
%30 & $10^{49}$ &  $1.79 \times 10^{29}$ & $4.67 \times 10^{27}$ \\
%30 & $10^{50}$ &  $1.10 \times 10^{30}$ & $2.79 \times 10^{27}$ \\
%30 & $10^{47}$ &  $2.7 \times 10^{27}$ & $1.8 \times 10^{27}$ \\
%30 & $10^{48}$ &  $2.4 \times 10^{28}$ & $7.6 \times 10^{27}$ \\
%30 & $10^{49}$ &  $1.8 \times 10^{29}$ & $4.7 \times 10^{27}$ \\
%30 & $10^{50}$ &  $1.1 \times 10^{30}$ & $2.8 \times 10^{27}$ \\
30 & $10^{47}$ &  $1.3 \times 10^{-6}$ & $8.5 \times 10^{-7}$ \\
30 & $10^{48}$ &  $1.2 \times 10^{-5}$ & $5.2 \times 10^{-6}$ \\
30 & $10^{49}$ &  $8.9 \times 10^{-5}$ & $4.7 \times 10^{-6}$ \\
30 & $10^{50}$ &  $5.4 \times 10^{-4}$ & $3.7 \times 10^{-6}$ \\
\enddata
\end{deluxetable*}

%・fig density velocity temp
%・outermost region is accelerated to relativistic speed
%・temperature  > Fe disintegrate, 
\subsection{Free neutron layers}
\begin{figure*}[h]
%\plotone{velo_rho.eps}
\plottwo{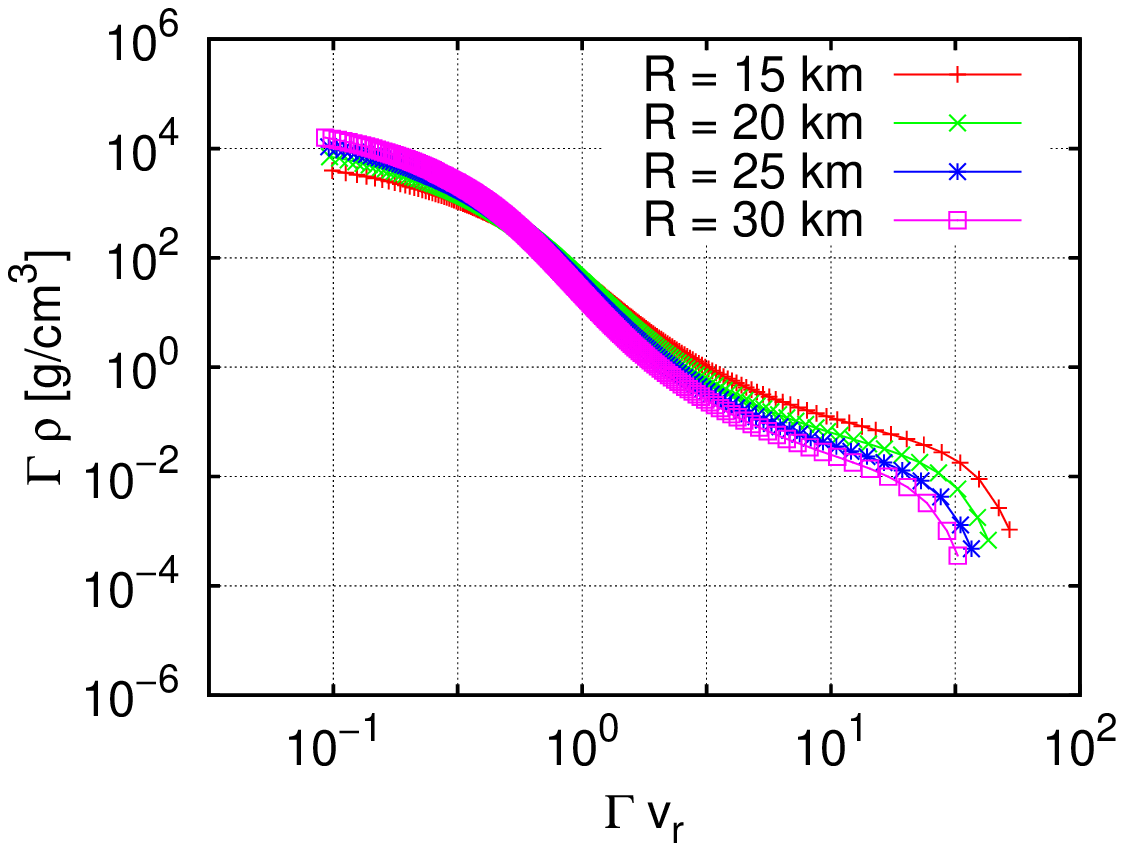}{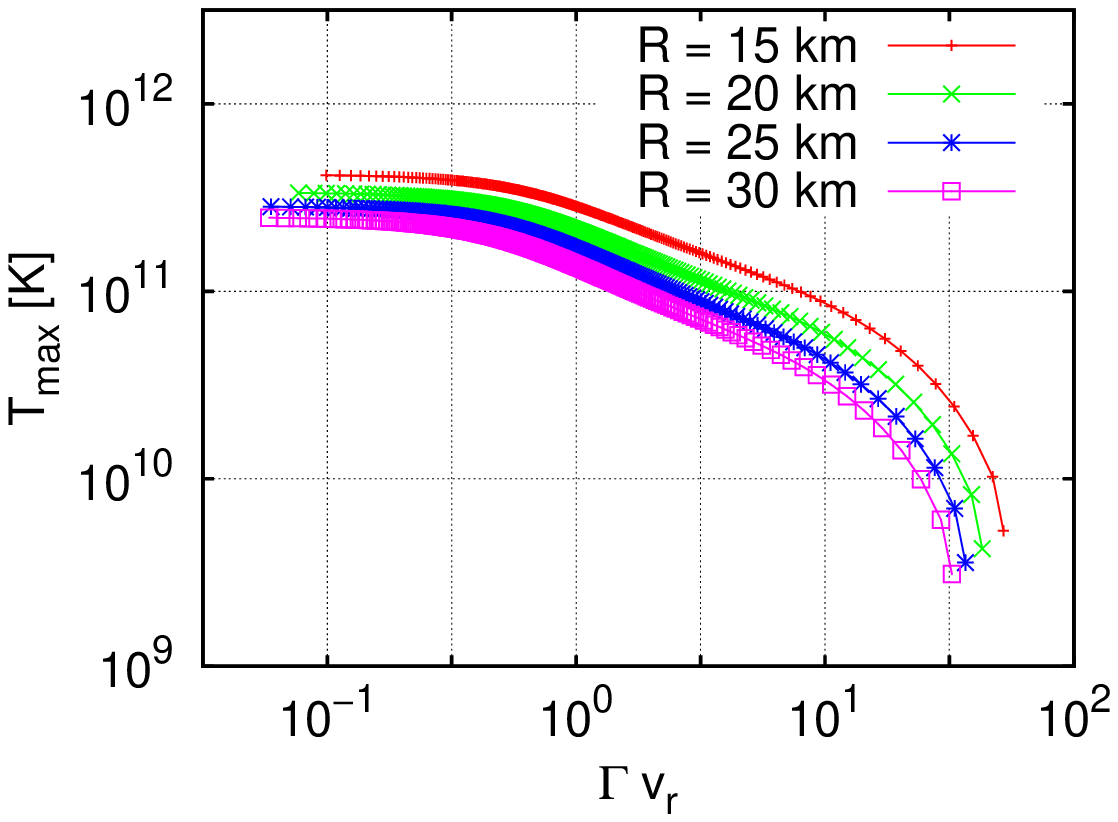}
\caption{Snapshot of the distribution of the mass density at $t = 1.8\times10^{-2}$ s (left panel), 
	and the distribution of the maximum temperature (right panel), with respect to the final velocity $\Gamma v_r$, for models with $E_{\rm f} = 10^{49}$ erg.}
\label{fig:velo_rho_temp}
\end{figure*}

%, which is calculated by mutiplying $X_{\rm n}$ by mass in the one computational cell and summing it all over the computational domain. 
Figure \ref{fig:velo_rho_temp} shows a snapshot of the distribution of the mass density $\Gamma\rho$ at $t=1.8\times10^{-2}$ s 
and a distribution of the maximum temperature of each fluid element 
with respect to the four-velocity $\Gamma v_r$ (in units of the speed of light) for models 
with $E_{\rm f} = 10^{49}$ erg. Note that these figures show only the ejected part of the envelope. 
%with parameters which can generate the largest amount of $M_n$; 
%that is, $R$ = 15 km and $E_{\rm f} = 10^{48}$ erg, $R$ = 20, 25, 30 km and $E_{\rm f} = 10^{49}$ erg.
%Although the innermost ejecta has non-relativistic speed, the outermost ejecta has relativistic speed with a few tens of Lorentz factor. Density in the outermost region is smaller than the innermost one by eight orders. 
%Figure \ref{fig:velo_temp} shows the final distribution of maximum temperature 
%during the all computational time at each computational cell. 
%%
%The distributions of the maximum temperature have peaks between $\Gamma v_r = 10^{-1}$ and 1. 
%This is a result of two competing effects. 
%A shock wave formed by a sudden injection of energy pushes matter in the outer region 
%with lower pressures inversely proportional to the volume swept by the shock. 
%At the same time, lower densities in the outer region reduce the fraction of the degeneracy pressure to the total pressure, 
%thus which leads to a higher temperature there. 
%After the degeneracy pressure at the shock becomes negligible, then the reduction of pressure at the shock directly results in the reduction of temperature. 
%This happens in the fluid elements between $\Gamma v_r = 10^{-1}$ and 1.
%%

\begin{figure}[h]
\begin{center}
\includegraphics[width=0.5\textwidth]{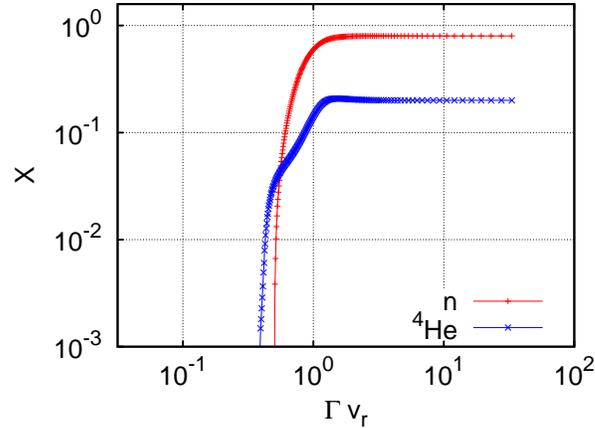}
\end{center}
\caption{Distributions of the mass fraction of free neutrons and $^4$He, with respect to the final velocity $\Gamma v_r$, for models 
	with $R = 25$ km and $E_{\rm f} = 10^{49}$ erg at $t =9.3\times10^{-2}$ s, with a temperature $< 10^8$ K in all the cells. }
\label{fig:velo_Xn}
\end{figure}

\begin{figure}[t]
\plottwo{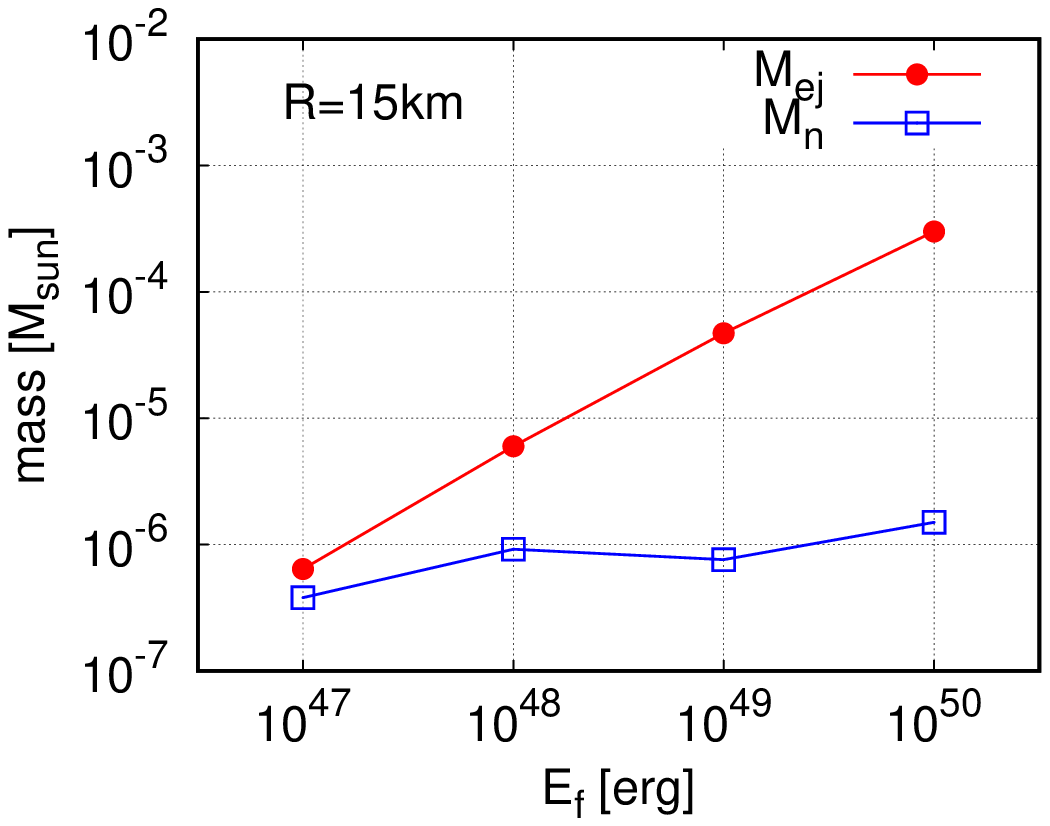}{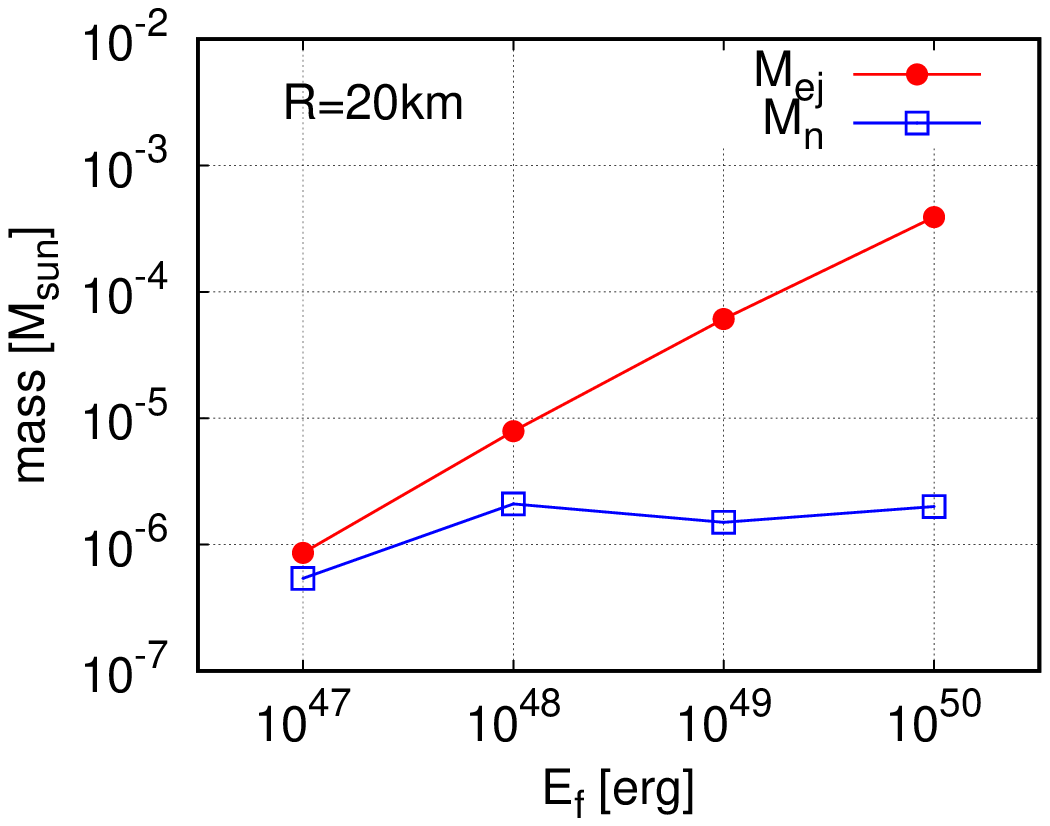}
\plottwo{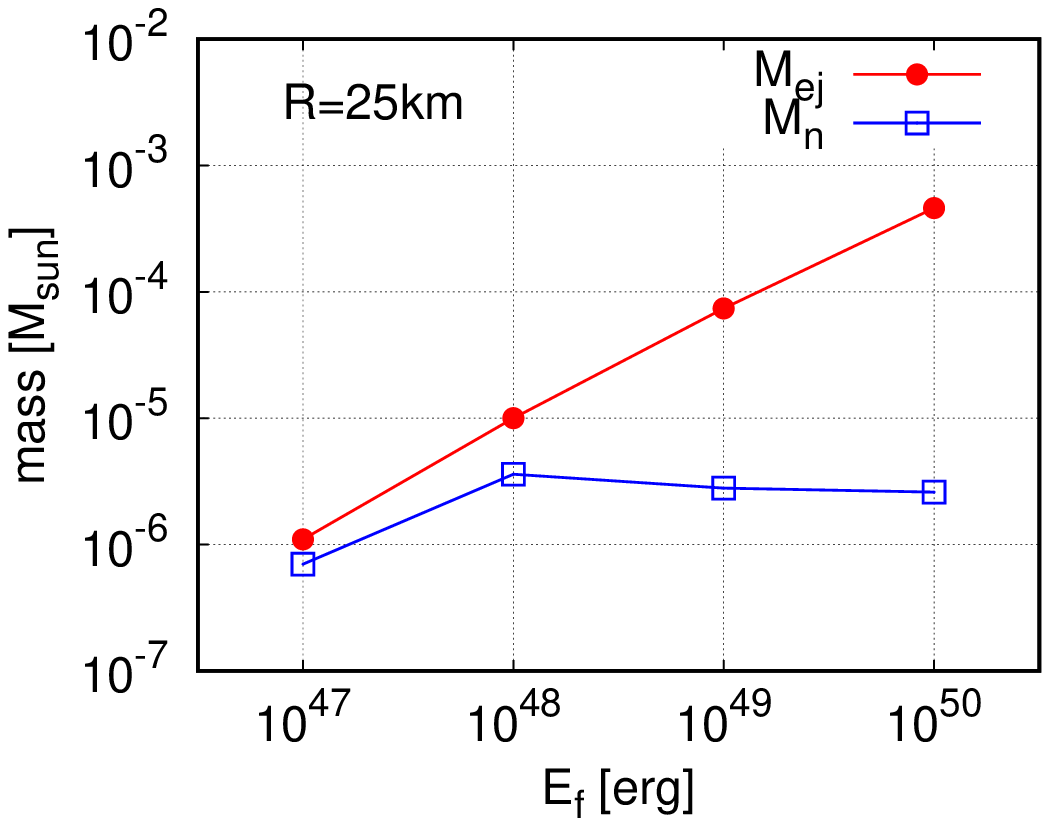}{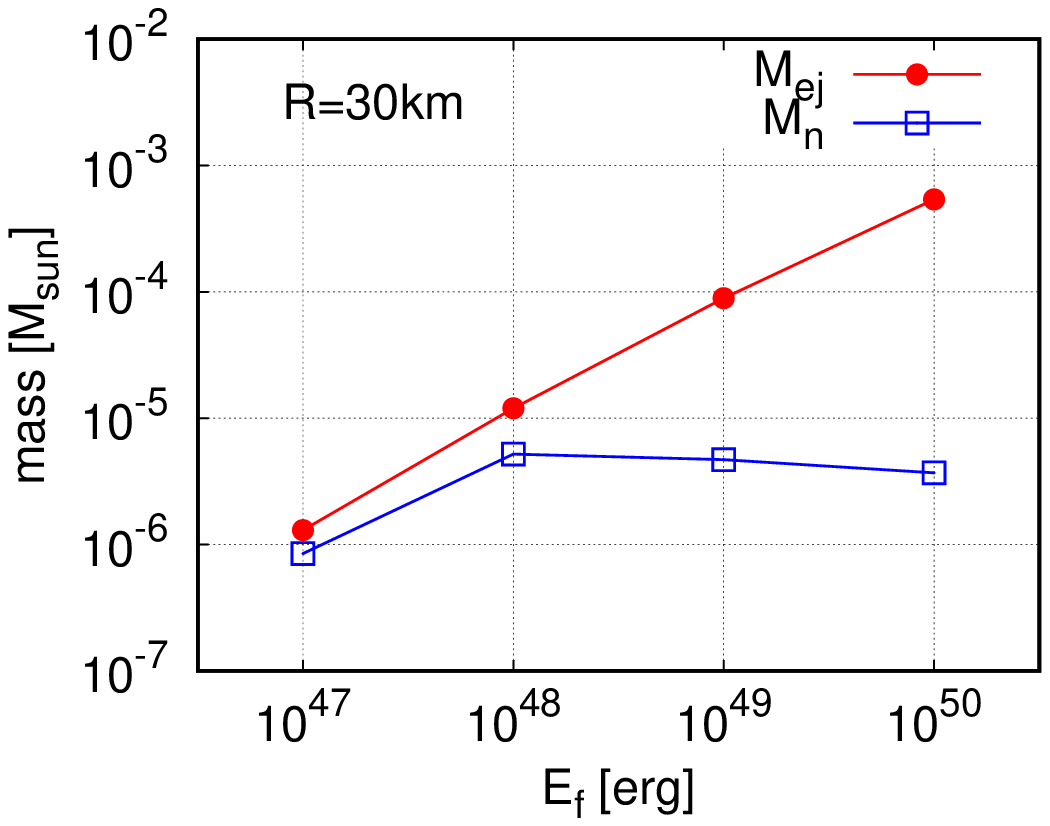}
\vspace{10mm}
\caption{Mass $M_{\rm ej}$ of ejecta and  the mass $M_{\rm n}$ of ejected free neutrons with respect to $E_{\rm f}$. 
	The top left panel shows models with $R=15$ km, the top right panel shows those with $R=20$ km, 
	the bottom left panel shows those with $R=25$ km, and the bottom right panel shows those with $R=30$ km.}
\label{fig:energy_mass}
\end{figure}

Figure \ref{fig:velo_Xn} shows the results of nucleosynthesis calculations when the temperature in all the cells becomes $< 10^8$~K.  
%The  maximum temperatures greater than  $10^{11}$ K significantly reduce $X_{\rm n}$ due to positron captures and convert most of the component into protons 
%($\Gamma v_r<0.3$ in Figure \ref{fig:velo_Xn}). 
%{\bf The maximum temperatures greater than $10^{11}$ K make the rates for electron and positron captures equal.} 
%{\bf In the inner region, density and temperature are high enough to synthesize the heavy nuclei ($\Gamma v_r < 0.6$ in Figure \ref{fig:velo_Xn}).}
%In the region where the maximum temperature less than $10^{11}$ K, some free neutrons prevents from positron captures. 
%{\bf In the region where the maximum temperature less than $10^{11}$ K, the rate for the electron capture becomes low compared to the positron capture.} 
As the temperature decreases down below $10^{10}$ K, the reactions p(n,~$\gamma$)d occur and 
the resultant deuterons quickly fuse to produce $^4$He. 
In the inner region, one after another, the residual neutrons are captured by the nuclei and produce heavy neutron-rich elements 
($\Gamma v_r < 0.6$ in Figure \ref{fig:velo_Xn}). 
%As the temperature decrease down below $10^{10}$ K, the reactions p(n, $\gamma$)d consume all neutron if $X_{\rm n}< 0.5$. 
%The deuterons quickly  fuse to produce $^4$He ($0.3<\Gamma v_r<0.9$ in Figure~\ref{fig:velo_Xn}). 
%If $X_{\rm n}>0.5$, then the same reactions p(n, $\gamma$)d consume all protons and 
%neutron capture reactions take place with the residual neutrons to produce heavy neutron rich nuclei ($\Gamma v_r\sim1$). 
%the residual neutrons are captured by the other nuclei one after another to produce heavy neutron rich elements ($\Gamma v_r\sim1$). 
%In the outer layer ($\Gamma v_r>0.6$), the lower density prohibits neutron captures. 
In the outer layer ($\Gamma v_r>0.6$), the lower density prohibits neutron captures. 
Consequently, the nucleosynthesis no longer occurs and the outermost layers contain free neutrons and $^4$He. 
In other words, the r-process freezes out in the outer layer. 
%In models with lower energies, the ejecta do not include the inner proton rich layer while models with higher energies increase the proton rich layer. 
%In models with lower energies, the ejecta do not include the inner proton rich layer while a model with a higher energy has a more extended proton rich layer. 
%{\bf Note that the final values of $Y_{\rm e}$ are shown in Figure \ref{fig:velo_Xn}.} 
Since the mass of the ejecta is concentrated in the inner region ($\Gamma v_r<1$) as indicated by the left panel of Figure \ref{fig:velo_rho_temp}, 
the free neutron layer occupies only a fraction of the total ejecta mass.
The ejecta masses $M_{\rm ej}$  and the ejected free neutron masses $M_{\rm n}$ in each model, 
with different $R$ and $E_{\rm f}$ are plotted in Figure~\ref{fig:energy_mass} 
and summarized in Table~\ref{tab:table}.
%So far, the initial $X_{\rm n}$ is determined by the beta equilibrium after the shock passage to examine the lower limit of the final $X_{\rm n}$. In addition, we have picked up the model from which we obtain the maximum yield of neutrons and change the initial fraction $X_{\rm n}$ of neutrons to 0.9 mimicking the beta equilibrium in a cool dense matter before the shock passage. Then we calculate the subsequent temporal change of the fraction solely due to positron captures. This assumption will give the maximum yield of neutrons among our models. The snapshot for the distribution of the mass fraction $X_{\rm n}$ is shown in the right panel of Figure \ref{fig:velo_Xn} and Table \ref{tab:table2} with $E_{\rm f} = 10^{49}$ erg. 
%in the cases that the mass $M_{\rm n}$ of free neutrons becomes maximum with each radius $R$ 
%($R$ = 15 km and $E_{\rm f} = 10^{48}$ erg, $R$ = 20, 25, 30 km and $E_{\rm f} = 10^{49}$ erg). The values of $M_{\rm n}$ become almost twice as those with the initial $X_{\rm n}$ determined by the beta equilibrium (eq. (\ref{eq:beta})). 
%The maximum $M_n$ of $4.72\times10^{28}$ g ($\sim 2.5\times10^{-5}M_{\odot}$) is obtained from a model with $R$ = 30 km and $E_{\rm f} = 10^{49}$ erg. 
%The maximum $M_n$ of $3.8\times10^{-6}\ M_{\odot}$ is obtained from a model with $R$ = 30 km and $E_{\rm f} = 10^{48}$ erg. 

The ejecta mass increases with increasing radius $R$ for the same $E_{\rm f}$ due to decreasing surface gravities. 
Though the ejecta mass also increases monotonically with increasing $E_{\rm f}$,  
a higher $E_{\rm f}$ leading to higher temperatures behind the shock wave shortens the timescale of the positron capture process through eq.~(\ref{eq:posi_cap}) and reduces $X_{\rm n}$. 
%{\bf In the models with the small radius $R$, the timescale for the expansion is short and the heavy nuclei synthesis does not actively occur. 
%So, the $M_{\rm n}$ becomes large with large $E_{\rm f}$ (that is, large $M_{\rm ej}$). 
%The fluctuation in the line for $M_{\rm n}$ in Figure~\ref{fig:energy_mass} is generated with the balance between the decrease of $M_{\rm n}$ by the positron captures 
%and the increase of $M_{\rm ej}$ as $E_{\rm f}$ increases.
In models with small $R$, positron captures monotonically decrease the fraction of free neutrons as a function of $E_{\rm f}$, 
while the faster expansion resulting from a larger $E_{\rm f}$ inhibits neutron captures by heavy nuclei. 
These two factors determine the behavior of $M_{\rm n}$ as a function of $E_{\rm f}$ in the top panels of Figure \ref{fig:energy_mass}. 
%In the models with the large $R$, the large timescale for the expansion leads the heavy nuclei synthesis.
%So, the value of $M_{\rm n}$ decreases with the increase of $E_{\rm f}$ through the positron capture and nucleosynthesis processes.}
%As a result, the  maximum mass $M_{\rm n}$ of free neutrons in the ejecta is obtained from models with low $E_{\rm f} = 10^{48}$ ($R$ = 15 km).  
%The maximum $M_{\rm n}$ is obtained for   $E_{\rm f} = 10^{49}$ erg ($R$ = 20, 25, 30 km) when  a low $E_{\rm f}(=10^{48}\,{\rm erg})$ decreases the initial $X_{\rm n}$ too much. 
In models with larger radii, the lower expansion rates do not significantly inhibit neutron capture by heavy nuclei 
even with a high $E_{\rm f}$. 
Thus, $M_{\rm n}$ does not increase in the models with $E_{\rm f} = 10^{50}$ erg in the bottom panels of Figure~\ref{fig:energy_mass}. 
As a result, 
%the  maximum mass $M_{\rm n}$ of free neutrons in the ejecta is obtained from models with $E_{\rm f} = 10^{48}$ erg. 
the maximum value of $M_{\rm n}$ over all of our models 
%is $\sim 2.49 \times 10^{28}$ g ($1.25\times10^{-5} M_{\odot}$) with $R$ = 30 km and $E_{\rm f} = 10^{49}$ erg. 
is $5.2\times10^{-6}\ M_{\odot}$, with $R$ = 30 km and $E_{\rm f} = 10^{48}$ erg. 

%The all ejecta is bound by gravitational force at first and a part of ejecta behind the shock wave is unbound at a certain time after expansion of ejecta, and the position of the firstly unbound ejecta corresponds to the peak of the distribution. The shock wave becomes weak as the shock wave propagates to the outer side; therefore, the temperature behind the shock wave decreases. The inner region than the peak is bound due to the gravitational force at first and unbound at the much later phase after the shock wave passes; therefore, the temperature of the region is smaller than the peak one. 
%Figure \ref{fig:velo_Xn} shows the final distribution of the free neutron surviving rate $X_{\rm n}$. 

%We considered naively that the outermost ejecta has the highest $X_{\rm n}$ because the temperature decreases rapidly with acceleration to the relativistic speed and the positron capture rate becomes small. However, the inner non-relativistic ejecta has the highest $X_{\rm n}$ and $M_n$. This shows it is more dominant effect to determine the highest $X_{\rm n}$ and $M_n$ that rather temperature is not so high around the shock wave after shock passing than the ejecta is cooled rapidly in the outermost region.

The emission from the outer ejecta composed of free neutrons is discussed in the next section. 

%%\startlongtable
%\begin{deluxetable*}{ccc}[t]
%\tablecaption{The masses of ejecta and ejected free neutrons for some models with an initial $X_{\rm n} = 0.9$ ($E_{f} = 10^{49}$ erg). \label{tab:table2}}
%\tablehead{
%\colhead{$R$} & \colhead{$M_{\rm ej}$} & \colhead{$M_{\rm n}$}\\
%\colhead{[km]} & \colhead{[g]} & \colhead{[g]}
%}
%%\colnumbers
%\startdata
%15 & $9.04 \times 10^{28}$ & $3.85 \times 10^{27}$ \\
%20 & $1.18 \times 10^{29}$ & $1.15 \times 10^{28}$ \\
%25 & $1.43 \times 10^{29}$ & $2.61 \times 10^{28}$ \\
%30 & $1.68 \times 10^{29}$ & $4.72 \times 10^{28}$ \\
%\enddata
%\end{deluxetable*}

%・/parameter: injection energy, delta r, shell mass
%・positron capture time scale eq.
%・free neutron mass fraction time derivative eq.
%・time derivative free neutron surviving rate for posi capture
%%%% If there is time
%・deuterium capture
%%%%
%・neutron rate distribution
%・after shock paths Fe is dissociated to neutron positron, He is also dissociated, calculation starts here

\section{conclusions and discussion} \label{sec:conclusion}
To examine the possibility of ejecting free neutrons  from the envelope of binary NSMs as previously claimed by  \citet{2015MNRAS.448..541J}, we have constructed simplified models in which spherically symmetric envelopes in hydrostatic equilibrium are expelled by strong shocks initiated by sudden injection of thermal energies.  To investigate what conditions maximize the yields of free neutrons, we have calculated 16 models by changing $R$ and $E_{\rm f}$. The mass of the envelope is set to be $10^{-3}\,M_\odot$ for all the models. 
Our results suggest that the ejecta mass is no more than a half of this initial envelope mass, 
even for an extremely large $E_{\rm f}$ of $10^{50}$ erg and $R$ of 30 km (see Table~\ref{tab:table}). 

Our series of simulations suggest that it is likely that an energy level of $10^{48}$ erg is preferable for free neutron ejection 
for the models with larger $R$. 
If the energy is smaller than this threshold, then the shock cannot eject enough amount of the envelope. 
If the energy is larger, too high temperatures behind the shock front convert free neutrons to protons by positron captures. 
The mass $M_{\rm n}$ of ejected free neutrons is maximized with $E_{\rm f} = 10^{48}$ erg 
and is of the order of $10^{-6}\,M_\odot$, even for a model with an extremely large $R$ = 30 km.  
We have obtained the values of $M_{\rm n}$ that are more than two orders of magnitude smaller than those in \citet{2015MNRAS.448..541J}. 
The distribution of $X_{\rm n}$ is %also different from the previous work 
%when the initial values are determined by the beta equilibrium immediately after the shock passage. 
%If the initial distribution is determined by the beta equilibrium before the shock arrival, 
%which should be more realistic, then the distributions of $X_{\rm n}$ become 
qualitatively consistent with the previous works, 
though the total amount of free neutrons is still significantly smaller.
%Here we should add one figure for a model with an initial Xn=0.9 and discuss the influence of initial conditions on the distribution of final Xn. 
%%%% from Tanaka comment
%This is because p(n, $\gamma$)d reaction is taken into account in this work 
%while it was not included in previous studies. 
%{\bf Moreover, in the previous work, the SPH simulation was performed with the small number of SPH particles in the outermost low-density region \citep{2015MNRAS.446.1115M}. 
%%So, the resultant values for $M_{\rm n}$ are not reliable.
%So, the difference for $M_{\rm n}$ between the previous study and our work might come from the resolution problem.} 
This indicates that free neutrons in the previous works originated from a different mechanism, 
such as the tidal debris and/or the neutrino-driven wind, 
while only the shock-heated ejecta are considered in this work. 

The initial mass fraction $X_{\rm n}$ is determined by assuming that 
the nuclei of heavy elements are disintegrated to free neutrons and protons 
because of shock heating, 
which is an endothermic reaction. 
The effects of the energy being consumed by the endothermic reaction are not considered in our models.
The energy density is calculated by $a_{\rm r} T^4 / \rho \sim 3 \times 10^{19}$ erg g$^{-1}$ at the shock front, 
with $T \sim 2.5 \times 10^{11}$. 
%in the case with $R = 25$ km and $E_{\rm f} = 10^{49}$ erg. 
Since the energy of the endothermic reaction for Fe is $\sim$9 MeV per nucleon, 
the energy density for the reaction is calculated by $9\ {\rm MeV} / m_{\rm u} \sim 8.6 \times 10^{18}$ erg g$^{-1}$. 
%where $m_{\rm u}$ is the atomic mass unit.
Thus, the energy is brought out by the endothermic reaction by a few tens of a percent. 
The shock speed and temperature at the shock front are decrease due to the reaction, 
and the positron capture process becomes unfavorable, resulting in a large amount of $X_{\rm n}$. 
Here, the largest amount of expelled energy by the endothermic reaction is considered for the most stable element, Fe; 
%because the composition of the merging neutron stars is unknown. 
thus the expelled energy might decrease for realistic compositions of the merging neutron stars. 

%%%% it is needed correction? Yes. TS revised it taking into account the results and added some comments on the previous work by Metzger+ 2015.
\citet{2015MNRAS.446.1115M} discussed the emission from ejecta composed of some free neutrons. 
They assumed that the opacity of the ejecta is due to lanthanide elements. 
Our models suggest that the ejecta eventually becomes composed of hydrogen, helium, 
and neutron-rich heavy elements after beta decays. 
In addition, the amount of free neutrons is much smaller than that in the previous studies. 
Nevertheless, depending on the viewing angle, one can observe emission from the ejecta in this study. 
Thus, the situation may change as follows. 
Suppose that the mass $M_{\rm ej}$ and energy of the ejecta are $10^{-5}\, M_{\odot}$ and $10^{48}$ erg, then the mean velocity of the ejecta becomes $\sim c/3$. 
The ejecta remains optically thick for $\sqrt{\kappa M_{\rm ej} / 4\pi}/(c/3) \sim 2.5\times10^3$ s as long as the matter is still fully ionized and the opacity is dominated by Thomson scattering, i.e., $\kappa \sim$ 0.4 cm$^2$ g$^{-1}$. 
Due to the small opacities, the characteristic timescale of the emission becomes shorter than previously expected. 
The time it takes for the diffusion velocity of photons to become comparable to the expansion velocity ($c/3$) is 
about $t_{\rm ch} \sim$ 1500 s. 

%The temperature in this layer can be roughly estimated as follows. 
The typical temperature and luminosity of the emission at this characteristic timescale can be estimated as follows. 
The beta decay of a neutron supplies an energy of 300 keV on a timescale of $\sim800$ s. 
Thus, the energy density at this time is given by 
$\epsilon_0 \sim 300$~keV~$\times(M_{\rm n}/m_{\rm n})/ [4\pi \times (800\,{\rm s} \times c/3)^3/3]\sim 10^6$~erg~cm$^{-3}
(M_{\rm n}/3.6\times10^{-6}\,M_\odot)$, 
which corresponds to a temperature of $\sim 10^5$ K, assuming radiation-dominated ejecta. 
Subsequent adiabatic expansion lowers this temperature 
and a typical temperature at $t_{\rm ch} \sim$ 1500 s is $\sim 5.3 \times 10^{4}$ K. 
%as long as the diffusion velocity of photons is smaller than the expansion velocity $c/3$. 
%During this phase lasting about 1,500 s, the luminosity $L$ is estimated as
The luminosity $L$ is estimated as 
\begin{equation}
%L\sim 4\pi (ct/3)^2 c\epsilon_0(800\ {\rm s}/t)^4\sim4.4\times10^{42}\ {\rm erg\ s^{-1}} \left(\frac{t}{1,500\,{\rm s}}\right)^{-2}\left(\frac{M_{\rm n}}{2.2\times10^{-6}\,M_\odot}\right).
L\sim 4\pi (ct/3)^2 v_{\rm diff}\epsilon_0(800\ {\rm s}/t)^4\sim 7.6\times10^{41}\ {\rm erg\ s^{-1} \left(\frac{t}{1500\,{\rm s}}\right)^{-2}\left(\frac{M_{\rm n}}{3.6\times10^{-6}\,M_\odot}\right)},
\end{equation}
where $v_{\rm diff}$ is the diffusion velocity when photons can diffuse out from the ejecta, which is $\sim c/9$.

According to the luminosity and temperature estimated above, 
the emission powered by decays of free neutrons mainly goes to ultraviolet wavelengths on a timescale of about 30 minutes. 
Assuming blackbody radiation, absolute magnitudes at 2000 and 2600 $\AA$ (UVW2 and UVW1 of {\it Swift}/UVOT) 
are about -13.55 and -13.25 mag, respectively (AB magnitude). 
These magnitudes correspond to the observed magnitudes of 19.45 and 19.75 mag at 40 Mpc, 
which are detectable with the {\it Swift} satellite if observations start immediately after the merger. 

The presence of hydrogen in the outermost layers may also suggest an emission similar to that of Type II-P supernovae. 
Recombinations start in the outer layers and the ejecta are expected to emit the blackbody radiation 
with a temperature of $\sim$6000 K, 
which corresponds to the recombination temperature of hydrogen in Type II-P supernovae 
\citep{1990ApJ...360..242S, 1993ApJ...414..712P, 2009ApJ...703.2205K}. 
From these values, the luminosity in this recombination phase can be roughly estimated as $6\times10^{38}$~erg~s$^{-1}(t/2.5\times10^3\,{\rm s})^2(T/6000\,{\rm K})^4$ 
at the final stage of the photospheric emission.

If the emission from such ejecta can be detected, 
some information on the dynamics of the merging event can be extracted 
because the spectrum must involve individual line features associated with transitions of hydrogen and helium. 
This is in contrast to dynamical ejecta composed of lanthanide elements 
that create too many inseparable lines in the spectrum and hide the information on dynamics.

%・/comparing to Metzger
%・effect of rotation force

\acknowledgments
We are grateful to the anonymous referee for providing helpful comments on this manuscript. 
This work was supported by JSPS KAKENHI grants No. 16H06341, 16K05287, 15H02082, 15H02075, 16H02183, 17H06357, 
and 17H06363.


%\vspace{5mm}
%\facilities{HST(STIS), Swift(XRT and UVOT), AAVSO, CTIO:1.3m,


%\software{astropy \citep{2013A&A...558A..33A},  
%          Cloudy \citep{2013RMxAA..49..137F}, 
%          SExtractor \citep{1996A&AS..117..393B}
%          }



\appendix

\section{Test for grid convergence}

\begin{figure}[t]
	\begin{center}
	%\plotone{velo_Xn_grid-500-1000_e-cap.eps}
 	\includegraphics[width=0.6\linewidth]{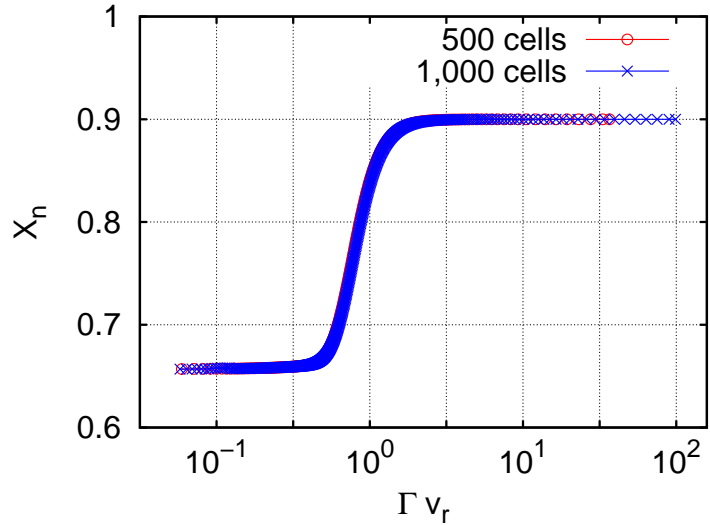}
	\caption{Comparison of distributions of $X_{\rm n}$  at the time $t=5\times10^{-3}$ s between computations with 500 and 1000 computational cells.}
	\label{fig:Xn_grid500_1000}
	\end{center}
\end{figure}

We have performed calculations for a model with $R$ = 25 km and $E_{\rm f} = 10^{49}$ erg 
using 500 and 1000 computational cells. 
%The initial value of $X_{\rm n}$ is calculated from beta equilibrium after shock passing. 
The calculation is initiated by increasing the thermal energy in the innermost 20 cells 
in the case with 1000 cells to generate a strong shock, 
which has the same volume as the innermost 10 cells in the case with 500 cells.   
Figure \ref{fig:Xn_grid500_1000} shows a comparison of the profiles of the $X_{\rm n}$ distributions resulting 
from calculations using 500 and 1000 computational cells. 
%Though the distributions deviate at $\Gamma v_r \gtrsim 10$, 
Though the distribution in the case with 1000 cells extends to a larger $\Gamma v_r$ region than that with 500 cells, 
the masses of ejecta and free neutrons are $7.4\times10^{-5}\ M_{\odot}$  and $2.8\times10^{-6}\ M_{\odot}$ 
for the model using 500 cells and $7.7\times10^{-5}\ M_{\odot}$ and $2.7\times10^{-6}\ M_{\odot}$ for the model with 1000 cells.
Therefore, the computations with 500 cells have already achieved enough convergence for our present purposes. 

{}



\end{document}